\documentclass[prb,aps,amsmath,amssymb,amsfonts,floatfix,superscriptaddress,
showpacs,twocolumn]{revtex4}
\usepackage{graphicx}
\usepackage[dvips]{color}
\usepackage{dcolumn}
\newcommand{\fig}[1]{Fig.~\ref{#1}}

\begin{document}

\title{Theoretical evidence for strong correlations and
incoherent metallic state in FeSe}

\author{Markus~Aichhorn}
\affiliation{Centre de Physique Th\'eorique, \'Ecole Polytechnique, CNRS,
91128 Palaiseau Cedex, France}
\author{Silke~Biermann}
\affiliation{Centre de Physique Th\'eorique, \'Ecole Polytechnique, CNRS,
91128 Palaiseau Cedex, France}
\affiliation{Japan Science and Technology Agency, CREST, Kawaguchi
  332-0012, Japan}
\author{Takashi~Miyake}
\affiliation{Japan Science and Technology Agency, CREST, Kawaguchi
  332-0012, Japan}
\affiliation{Research Institute for Computational Sciences, AIST, Tsukuba 305-8568, Japan}
\affiliation{Japan Science and Technology Agency, TRIP, Kawaguchi
332-0012, Japan}
\author{Antoine~Georges}
\affiliation{Centre de Physique Th\'eorique, \'Ecole Polytechnique, CNRS,
91128 Palaiseau Cedex, France}
\affiliation{Japan Science and Technology Agency, CREST, Kawaguchi
  332-0012, Japan}
\affiliation{Coll\`ege de France, 11 place Marcelin Berthelot, 75005 Paris}
\author{Masatoshi~Imada}
\affiliation{Japan Science and Technology Agency, CREST, Kawaguchi
  332-0012, Japan}
\affiliation{Japan Science and Technology Agency, TRIP, Kawaguchi
332-0012, Japan}
\affiliation{Department of Applied Physics, University of Tokyo, %7-3-1 Hongo,
  Bunkyo-ku, Tokyo 113-8656, Japan}

\begin{abstract}
The role of electronic Coulomb correlations in
iron-based superconductors is an important open question.
We provide theoretical evidence for strong correlation effects in
FeSe, based on dynamical mean field calculations. 
Our {\it ab initio} spectral properties first demonstrate the existence of a lower Hubbard band. 
Moreover, together with
significant orbital-dependent mass enhancements, we find that the normal state
is a bad metal over an extended temperature range, implying a non-Fermi liquid
due to formation of local moments.
Predictions for angle-resolved photoemission spectroscopy are made.
\end{abstract}

\maketitle

\section{Introduction}

The discovery of high-temperature superconductivity in iron-based
compounds triggered a tremendous amount of experimental and
theoretical research.
In general terms one
can divide these materials into two classes. The first is based on
iron-pnictogen FePn$^{-1}$ planes, which have to be intercalated with
some spacer atoms.
The second class is built up by
charge-neutral iron-chalcogenide FeCn layers, where Cn can be S, Se, or Te.
Superconductivity in this `11' family 
was first reported by Hsu {\em et al.},\cite{hsu2008} with a
transition temperature of $T_c\approx 8$\,K for $\alpha$-FeSe$_{0.85}$.
The properties of this
material can be further modified by
pressure,\cite{mizguchi2008APL,garbarino2009EPL,margadonna2009PRB}
excess Fe or Se 
deficiencies. Another interesting route is alloying FeSe and FeTe
yielding the family of FeSe$_x$Te$_{1-x}$ superconductors\cite{yeh2008}.

In contrast to the
high-temperature superconducting cuprates,
whose parent compounds are Mott insulators, the question about
the strength of correlations in the iron-based superconductors
is not settled. 
FeSe$_{1-x}$ has been studied with angular-integrated photoemission (PES) in
Refs.~[\onlinecite{yoshida2009JPSJ},\onlinecite{yamasaki2009u}], while
angular-resolved photoemission studies (ARPES) have been performed on
Fe$_{1+x}$Te\cite{xia2009} and FeSe$_x$Te$_{1-x}$
(with $x=0.3$ in [\onlinecite{nakayama2009u}] and $x=0.42$ in
[\onlinecite{tamai2010PRL}]).
For all these compounds, these 
experiments reveal a significant
iron bandwidth narrowing, by a factor in excess of $\sim 2$.
Interestingly, recent ARPES experiments on FeSe$_x$Te$_{1-x}$
with $x=0.42$\cite{tamai2010PRL} report large mass enhancements 
deduced from the low-energy Fermi velocities,
ranging from $6$ to $23$,
while a smaller enhancement (in the range $2-3$) was reported
for Fe$_{1+x}$Te\cite{xia2009}.
Specific heat measurements\cite{sales2009PRB} for
FeTe$_{0.67}$Se$_{0.33}$ gave $\gamma= 39$\,mJ/molK$^2$,
whereas for FeSe$_{0.88}$ a smaller value $\gamma=9.17$\,mJ/molK$^2$ was reported.\cite{hsu2008}

Band structure calculations based on density-functional theory (DFT)
of FeSe and FeTe have shown\cite{subedi2008PRB} that the one-electron
band structure of these materials is 
similar to the other iron-based
superconductors, as regards the Fe-$d$ states around Fermi level.
However, a recent first-principles calculation of the screened Coulomb
interactions gave significantly higher values for the 11
compounds.\cite{miyake2010} Dynamical mean-field (DMFT) calculations
at $T=0$ 
-- albeit within the iterative perturbation theory --
could reproduce some aspects of the experimental results, by taking the 
interaction as a parameter.\cite{craco_2009u}
Given the experimental controversies, 
further insight from {\rm ab initio} theoretical predictions, in
comparison with experiments, is needed in order
to characterize the role of
correlations for the 11 family relative to other iron superconductors 
(e.g. the 1111 family as LaFeAsO).

In this article, we investigate the correlation effects
on the electronic structure of $\alpha$-FeSe in the framework of
DMFT, using numerically exact Monte Carlo simulations and
taking into account the full rotationally invariant Coulomb
interactions, evaluated from first principles.
We find that this material
displays clear evidence of strong correlations. 
First, our results demonstrate that
the satellite feature observed in PES at a binding energy
of about $-2$ eV\cite{yoshida2009JPSJ,yamasaki2009u} should be
interpreted as a lower Hubbard band of iron origin. 
Second, we find that three
of the iron orbitals are characterized by a rather low
quasiparticle weight, and short quasiparticle lifetimes. These features
indicate deviation from conventional Fermi liquid behavior possibly
associated with the formation of local moments
controlled by the Hund coupling. 
The possible occurence of local moments in multi-band systems was discussed 
in a model calculation\cite{werner3band2008} and
for 1111 pnictides\cite{haule2,ishida2010PRB} 
for parametrized interaction strengths.
The situation in the 1111 is subtle. However, taking the interaction
strenght calculated from first priciples, LaFeAsO shows moderate
correlations.\cite{aichhorn2009} Here, on the contrary, our present
calculations with {\rm ab initio} interactions strongly suggest that
correlation effects are most clearly revealed in the 11 family.
We make theoretical predictions for the ARPES 
spectrum of FeSe. Significant deviations from the DFT band-structure are
found, which do not simply amount to an overall bandwidth narrowing.
A downward shift of the hole-like bands near the $\Gamma$-point and an upward shift
of the electron-like bands near the M-point are found. Although ARPES
has only been
reported at this stage for FeSe$_x$Te$_{1-x}$ , this
is consistent with the
trends reported for these alloys.\cite{tamai2010PRL,nakayama2009u}

\section{Methods}

Our calculations use the recent implementation of the combined DFT-DMFT
method in a full-potential augmented plane wave electronic-structure 
framework\cite{aichhorn2009} based on the Wien2k package.\cite{Wien2k}
Localized Wannier-like orbitals are constructed from an energy window comprising
the Fe-$3d$ bands and the Se-$4p$ bands, calculated in the local-density
approximation (LDA).
We use the tetragonal crystal structure, space group $P4/nmm$, as
reported by Margadonna {\em et al.},\cite{margadonna2008} using the
experimental value for the Se position in the unit cell.
A many-body self-energy, computed from DMFT using a strong-coupling continuous-time
quantum Monte Carlo algorithm,\cite{werner_ctqmc} is applied to the subspace spanned
by the Fe-$3d$ orbitals.
The matrix of Coulomb interaction parameters is calculated from first
principles using the constrained random-phase approximation (cRPA),\cite{ferdi_cRPA}
as applied to the iron-based superconductors in
Refs.~[\onlinecite{miyake1,miyake2010}]. All screening transitions are included,
except the ones within the Fe-$3d$ manifold. In the notation of Refs.~[\onlinecite{miyake1,miyake2010}] this 
corresponds to the so-called $d$-$dp$ construction.
In order to avoid an orbital dependent double-counting correction, we
do not use the calculated interaction matrices directly. Instead, we
follow the procedure described in Ref.~[\onlinecite{aichhorn2009}] and
use the orbitally-averaged Coulomb interactions.
For FeSe, this yields a local Coulomb integral
$U=F_0=4.06$\,eV and a Hund's coupling $J=0.91$\,eV. 
This is to be compared with the significantly smaller values
$U\simeq 2.7$\,eV, $J\simeq 0.8$\,eV for LaFeAsO.\cite{aichhorn2009,miyake2010}
Please not that 
the definition of $J$ is used here in terms Slater integrals, and hence
differs from the definition of $J$ in Refs.~[\onlinecite{miyake1,miyake2010}].
We will show that the inclusion of the full rotationally-invariant Hund's
coupling (including spin-flip and pair-hopping) is crucial for FeSe.
If not otherwise specified, all calculations were performed at a temperature $T=290$\,K 
using the fully-localized-limit double counting.\cite{Ylvisaker_LSDA+U_2009_prb}
Spectra were obtained from the imaginary-frequency Monte Carlo data using the
stochastic Maximum Entropy method.\cite{beach_ME}

\begin{figure}[t]
  \centering
  \includegraphics[width=0.8\columnwidth]{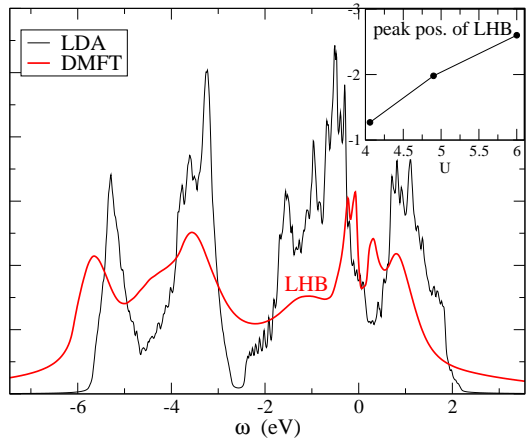}
  \caption{\label{DOStot}%
    (Color online) Comparison of the total LDA DOS (black) to the
    spectral function
    obtained with DMFT (red). 'LHB' denotes the lower
    Hubbard band. The inset shows the evolution of the LHB as a function
    of $U$, with $J$ scaled accordingly ($J=0.9$, 1.1, 1.3, respectively.)
  }
\end{figure}

\begin{figure}[t]
  \centering
  \includegraphics[width=0.8\columnwidth]{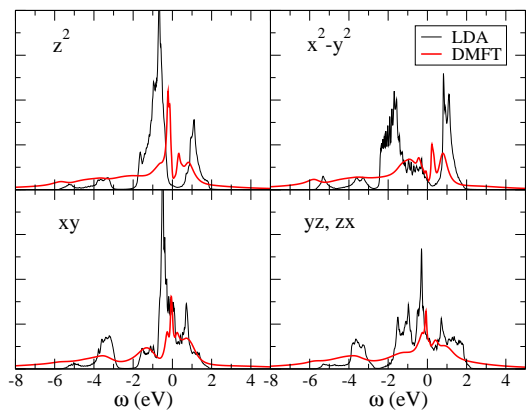}
  \caption{\label{partDOS}%
    (Color online) Orbital-resolved comparison of the density of states between the LDA
    (thin black) and DMFT (thick, red) results. The orbitals in the
    bottom row show strongest correlations. Note the LHB in the
    $d_{xy}$ orbital.
  }
\end{figure}

\section{Results}

In Fig.~\ref{DOStot}, we compare the total
momentum-integrated
spectral function obtained within LDA and within DMFT.
A substantial bandwidth reduction is obtained for the
Fe-$d$ states near the Fermi energy.
Furthermore, the Se-$p$ ligand states are
shifted to larger binding energies as compared to LDA.
In addition, there is a 
structure appearing at a binding energy between $-1$ and $-2$\,eV.
In order to identify the physical nature of this peak, we performed
calculations for larger interactions. For computational efficiency,
the calculations for varied $U$ and $J$ were done here with density-density
interactions only. Non density-density terms affect mostly the quasiparticle
states and little the LHB. %\cite{footnote1}.
It is obvious from the inset of \fig{DOStot} that the peak shifts towards larger
binding energies with increasing $U$, which is the expected behavior for a lower Hubbard band (LHB),
and distinguishes it from a low-energy quasi-particle excitation.
The existence of a lower Hubbard band is nicely consistent with
experimental PES results. Independent measurements on FeSe\cite{yoshida2009JPSJ,yamasaki2009u} revealed a broad feature at a
binding energy of about about $-2$\,eV. It was shown in Ref.~[\onlinecite{yoshida2009JPSJ}]
that the photon-energy dependence of this peak
(labeled `B' in Fig.~1 of Ref.~[\onlinecite{yoshida2009JPSJ}]) 
indicates that it is of Fe-$d$ origin.
Given our theoretical results, we propose that the feature observed in
Refs.~[\onlinecite{yoshida2009JPSJ,yamasaki2009u}] in this range of binding
energies should be
interpreted as a lower Hubbard band.
To our knowledge, this is indeed the first observation of a lower Hubbard band
in iron-based pnictide and chalcogenide superconductors. Indeed, the
shoulder at about -1.5 eV observed in PES for LaFeAsO  
in Ref.~\onlinecite{Malaeb1} can be explained from band theory. %\cite{footnote3}.
We also note that the position of the Se-$p$ bands 
agrees well with the
experimental peak positions of $-6$ and $-4$\,eV, corresponding to peaks 'C' and 'D' in
Fig.~1 of Ref.~[\onlinecite{yoshida2009JPSJ}].

Our results reveal a marked orbital dependence of the
correlation effects, with significantly stronger correlations for the
three orbitals $d_{xy}$ and $d_{xz,yz}$, while the $d_{z^2}$ and $d_{x^2-y^2}$ orbitals
display weaker correlations.
Our conventions for the $(x,y)$ axis are rotated by 45 degrees as
compared to the crystallographic axis, so that the $d_{xy}$ orbital in our definition
is the one pointing from Fe to Se. The three orbitals displaying stronger correlations are
thus the ones which form the peak of the LDA-DOS (and hence have higher weight at the
Fermi level), while the two other ones display a `pseudo-gap' in the LDA DOS.
In \fig{partDOS}, we plot the spectral functions of each Fe-$d$ orbital, calculated for
the same parameters as before. It is apparent that the LHB discussed above is visible
mostly in the $d_{xy}$ and $d_{xz,yz}$ partial DOS.
We have also calculated
$Z_m = (1-\frac{\partial \Sigma}{\partial i \omega})^{-1}|_{i\omega \to 0}$, which
-- in a Fermi liquid (see however below) -- can be
interpreted as the quasiparticle
weight, and the inverse
single-particle lifetime ${\rm Im}\Sigma_m(i0^+)$ of each orbital.
We find that the $d_{z^2}$ and $d_{x^2-y^2}$ orbitals yield larger
quasiparticle weights $Z_{z^2}=0.38$, $Z_{x^2-y^2}=0.47$ and
longer lifetimes (corresponding to $-{\rm Im}\Sigma(i0^+)$ of order $0.04$\,eV at
$T=290$\,K).
In contrast, a linear fit to the slope of the self-energy for
$d_{xy,xz,yz}$ orbitals would yield lower
values of the weights $Z_{xy}=0.20$, and $Z_{xz,yz}=0.28$. However, the rather
short lifetimes (corresponding to $-{\rm Im}\Sigma(i0^+)_{xy}\simeq 0.19$\,eV and
$-{\rm Im}\Sigma(i0^+)_{xz/yz}\simeq 0.08$\,eV) shed serious doubt on the validity of a
coherent quasiparticle interpretation for those orbitals.
A further decrease of the
temperature from $T=290$\,K to $T=190$\,K does not reduce
%\ins{
$\Sigma(i0^+)$. In order to identify coherent quasi particles the
  temperature has to be lower than the width of the quasi-particle,
  i.e. $T<Z{\rm Im}\Sigma(i0^+)$. Here $Z{\rm Im}\Sigma(i0^+)$ for the
  $d_{xy}$ orbital corresponds to 440\,K, which is higher than 290 or
  190\,K and, hence, no coherent quasi-particle can be expected.
We checked that changing the double counting to ``around-mean-field''\cite{Ylvisaker_LSDA+U_2009_prb}
even increases the incoherence ($-{\rm
  Im}\Sigma(i0^+)_{xz,yz}\approx 0.4$). %}
We elaborate in more detail on this unconventional metallic state further at the
end of this paper. 
The stronger degree of correlations in FeSe, compared to LaFeAsO,
\cite{aichhorn2009}, manifests itself also in the partial charges of the
Fe-$d$ electrons. Due to electronic correlations, the electron charge
diminishes from 6.37 in LDA to 6.07 within LDA+DMFT,
which should be compared to the values for LaFeAsO of 6.40 (LDA) and
6.28 (LDA+DMFT). Please note that these changes do {\em not}
correspond to any doping effect, since the total electron count of
the crystal is still integer. It is merely a redistribution of charges
due to changes in the hybridisation between the Fe and the
ligand atoms.

\begin{figure}[t]
  \centering
  \includegraphics[width=0.75\columnwidth]{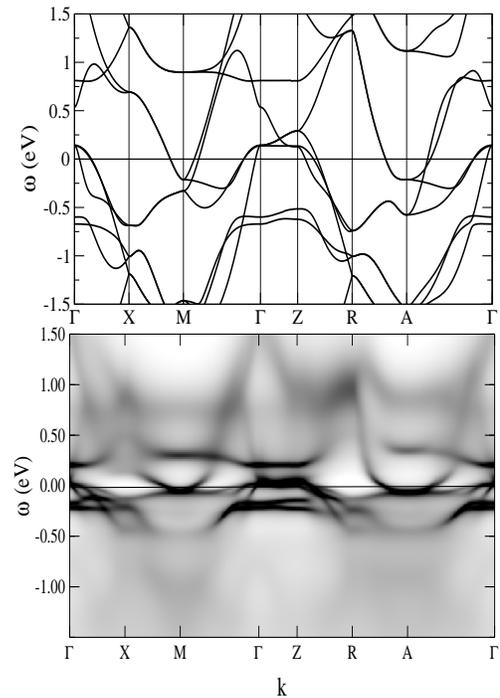}
  \caption{\label{Akwtot}%
    Comparison of the momentum-resolved spectral function of DFT (top) to the
    DFT+DMFT results (bottom).
  }
\end{figure}

\begin{figure}[t]
  \centering
  \includegraphics[width=0.8\columnwidth]{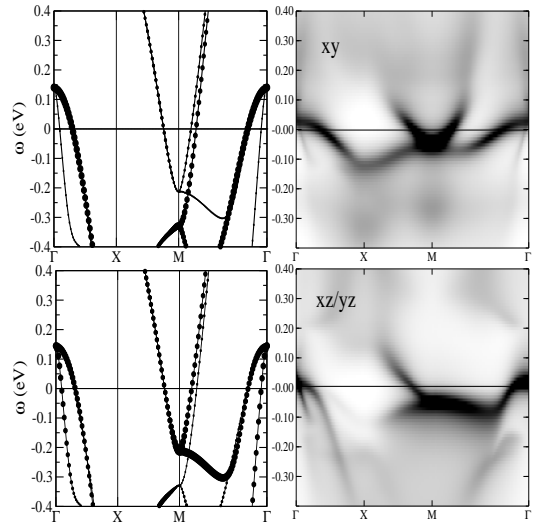}
  \caption{\label{Akwproj}%
    Comparison of the low-energy spectral function, projected to
    orbital character, of DFT (left) to the DFT+DMFT results
    (right). Top row: $d_{xy}$ orbital. Bottom row: Degenerate
    $d_{xz}$,$d_{yz}$ orbitals.
  }
\end{figure}
We have also calculated the momentum-resolved spectral function along high-symmetry
directions, displayed in \fig{Akwtot} and compared to the LDA bandstructure.
These results can be viewed as predictions for future ARPES experiments on
pure FeSe. They can also be compared qualitatively with available ARPES
data,\cite{tamai2010PRL,nakayama2009u} bearing in mind however that those actually concern the
Te-substituted compound FeSe$_x$Te$_{1-x}$.
The overall renormalization of the bands and the band-width reduction
are apparent.
The inverse lifetime ($\rm{Im}\Sigma$) increases rapidly with increasing frequency, which
results in rather broad structures below -0.3\,eV.
The DMFT quasiparticle bandstructure cannot be deduced from a mere rescaling of the
LDA bandstructure. The strong orbital and energy dependence of the self-energy renormalizes the
effective crystal-field splitting between the orbitals, resulting in differential shifts of
the quasiparticle bands near the Fermi level.
An interesting effect is seen for example near the $\Gamma$ point.
In LDA, one can see essentially only two dispersing hole-like bands,
the outermost band being quasi-degenerate.
By introducing correlations, this degeneracy is lifted,
and we can see three hole-like excitations around the $\Gamma$
point. This is qualitatively consistent with ARPES.\cite{tamai2010PRL}
We also observe that the hole pockets near the $\Gamma$-point are pushed downwards in
energy by correlation effects, while the electron-like ones near the M-point are
pushed upwards, in agreement with the ARPES spectra presented in Refs.~[\onlinecite{nakayama2009u, tamai2010PRL}]
The orbital character of the LDA+DMFT bands, see \fig{Akwproj},
reveals that the outermost band around $\Gamma$ and the electron
pocket around M are dominantly of $d_{xy}$ character, which displays the
largest effective mass in our calculations, see above. This is again
in qualitative agreement with ARPES, where the heaviest masses were
associated to the outermost hole pocket and the electron pocket.
Furthermore, the broader feature at around -0.3\,eV is of $d_{z^2}$
character, consistent with experiment.

Within the DMFT approach used in the present work, the effective mass
enhancement of electrons in a given orbital $m$ is related to
the inverse of the weight $Z_m$  calculated above.
This suggests effective mass enhancements in the range $\sim 2$ for the
$x^2-y^2$ orbital to $\sim 5$ for the $xy$ one.
Direct comparison of these values to experiments is difficult in the absence of
ARPES data for FeSe.
ARPES measurements on FeSe$_x$Te$_{1-x}$\cite{nakayama2009u,tamai2010PRL} have been
interpreted as yielding very large effective masses ($m^*/m_{\rm{band}}$ between 6
and 23)\cite{tamai2010PRL}.
This is qualitatively consistent with the much larger specific heat
coefficient reported for these alloys,\cite{sales2009PRB}
in comparison to FeSe.\cite{hsu2008}
It should also be kept in mind that the precise extraction of effective masses
(renormalized low-energy Fermi velocities) from ARPES data by comparison to
the DFT band structure is subject to rather large uncertainties, especially
in view of the orbital-dependent shifts of the bands induced by
correlations
(see above).

\begin{figure}[t]
  \centering
  \includegraphics[width=0.8\columnwidth]{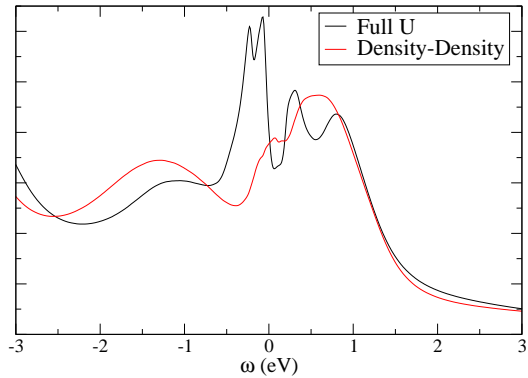}
  \caption{\label{compHund}%
    Difference of the total DOS when calculated using the full 4-index
    $U$-matrix (black) and density-density interactions (red). For both
    calculations: $U=4.06$\,eV, $J=0.91$\,eV.
  }
\end{figure}

Considering the full rotationally-invariant
Hund's coupling\cite{pruschke2005,haule2,ishida2010PRB} is crucial to
describe properly the low-energy physics of FeSe. 
Including only density-density interactions
drastically suppresses the quasi-particle at low-energies (see
\fig{compHund}), rendering the 
system much more incoherent for the same value of parameters, as 
reflected in high
scattering rates $-{\rm Im}\Sigma(i\omega^+)$ in the range $0.5-1.0$.
In this case also, lowering the temperature (here, down to
$T=50$\,K) does not reduce the scattering rates.
An increase of the Coulomb parameters by even 50\% does not
drive the system to a Mott-insulator: at $U=6.0$\,eV, $J=1.3$\,eV
there is still finite spectral
weight at the Fermi level. We increased $U$
further to $U=10$\,eV which finally leads to insulating behavior.
Taken together, our results for both rotationally invariant and
density-density interactions reveal that there is region of the $(U,J)$
parameter space, where the system is in a state which is neither a Fermi-liquid, nor
a Mott insulator.
This region in parameter space looks similar 
to the selective localisation found in Refs.~[\onlinecite{medici2009,craco_2009u}],
although we do not find one Mott-localised orbital, but instead three
incoherent orbitals. 
This is also consistent with recent
reports within DFT+DMFT calculations based on exact diagonalisation
for LaFeAsO\cite{ishida2010PRB} and FeSe.\cite{liebsch2010u}

This suggests that FeSe is just beyond the transition
from the Fermi-liquid phase to an intermediate bad-metallic phase,
and could be a realization of the 'spin-freezing'
scenario.\cite{werner3band2008,haule2} Indeed, the response to an 
external magnetic field increases significantly, when this phase is
entered (for instance by tuning $J$), whichs reveals the formation of
localised moments.
This appears to be consistent with the rather large values of the resistivity
measured even on single-crystal samples\cite{braithwaite2009} and with NMR\cite{imai_PRL2009}
and neutron scattering experiments\cite{lumsden_NaturePhysics2010} suggestive of strong
spin fluctuations.

\section{Conclusions}

We have presented strong theoretical and first-principles evidence for
enhanced electronic correlation effects in FeSe.
We have identified a Hubbard satellite in the spectral function, large orbital-dependent
mass enhancements ranging from 2 to 5, and -importantly -
large quasi-particle damping effects that may render a Fermi
liquid description of the normal state of FeSe invalid
over a wide temperature range.
Concerning the momentum-resolved spectral function, we find very good
agreement with recent ARPES measurments. In addition to the mass
enhancement, we found that a correlation induced crystal-field
splitting is important to describe size and position of the hole and
electron pockets at the $\Gamma$ and M point of the Brilloiun zone,
respectively. As argued in Ref.~[\onlinecite{nakayama2009u}], a simple scaling
of the LDA bands is not enough to account for the measured electronic
structure. 
The present orbital-sensitive non-Fermi liquid behavior further indicated by their sensitivity to the Hund's coupling 
may serve for understanding the interplay between the spin-charge-orbital dynamics and
superconductivity in iron superconductors.

\acknowledgments
MA, SB, and AG acknowledge discussions with 
V. Vildosola, L. Pourovskii, M. Ferrero, O. Parcollet,
A. Liebsch, H. Ding, L. Craco, 
L. de' Medici and D. van der Marel.
TM and MI thank discussions with R. Arita, K. Nakamura and T. Misawa. 
This research was supported in part by the Agence Nationale
de la Recherche under grant CORRELMAT,
of IDRIS/GENCI (project 101393), the National Science Foundation
under Grant No. PHY05-51164, and (MA) the
Austrian Science Fund (FWF) under grant J2760.
The hospitality of the Kavli Institute for Theoretical
Physics is gratefully acknowledged.


\begin{thebibliography}{32}
\expandafter\ifx\csname natexlab\endcsname\relax\def\natexlab#1{#1}\fi
\expandafter\ifx\csname bibnamefont\endcsname\relax
  \def\bibnamefont#1{#1}\fi
\expandafter\ifx\csname bibfnamefont\endcsname\relax
  \def\bibfnamefont#1{#1}\fi
\expandafter\ifx\csname citenamefont\endcsname\relax
  \def\citenamefont#1{#1}\fi
\expandafter\ifx\csname url\endcsname\relax
  \def\url#1{\texttt{#1}}\fi
\expandafter\ifx\csname urlprefix\endcsname\relax\def\urlprefix{URL }\fi
\providecommand{\bibinfo}[2]{#2}
\providecommand{\eprint}[2][]{\url{#2}}

\bibitem[{\citenamefont{Hsu et~al.}(2008)\citenamefont{Hsu, Luo, Yeh, Chen,
  Huang, Wu, Lee, Huang, Chu, Yan et~al.}}]{hsu2008}
\bibinfo{author}{\bibfnamefont{F.-C.} \bibnamefont{Hsu}},
  \bibinfo{author}{\bibfnamefont{J.-Y.} \bibnamefont{Luo}},
  \bibinfo{author}{\bibfnamefont{K.-W.} \bibnamefont{Yeh}},
  \bibinfo{author}{\bibfnamefont{T.-K.} \bibnamefont{Chen}},
  \bibinfo{author}{\bibfnamefont{T.-W.} \bibnamefont{Huang}},
  \bibinfo{author}{\bibfnamefont{P.~M.} \bibnamefont{Wu}},
  \bibinfo{author}{\bibfnamefont{Y.-C.} \bibnamefont{Lee}},
  \bibinfo{author}{\bibfnamefont{Y.-L.} \bibnamefont{Huang}},
  \bibinfo{author}{\bibfnamefont{Y.-Y.} \bibnamefont{Chu}},
  \bibinfo{author}{\bibfnamefont{D.-C.} \bibnamefont{Yan}},
  \bibnamefont{et~al.}, \bibinfo{journal}{Proc. Nat. Acad. Sci. U.S.A.}
  \textbf{\bibinfo{volume}{105}}, \bibinfo{pages}{14262}
  (\bibinfo{year}{2008}).

\bibitem[{\citenamefont{{Mizuguchi} et~al.}(2008)\citenamefont{{Mizuguchi},
  {Tomioka}, {Tsuda}, {Yamaguchi}, and {Takano}}}]{mizguchi2008APL}
\bibinfo{author}{\bibfnamefont{Y.}~\bibnamefont{{Mizuguchi}}},
  \bibinfo{author}{\bibfnamefont{F.}~\bibnamefont{{Tomioka}}},
  \bibinfo{author}{\bibfnamefont{S.}~\bibnamefont{{Tsuda}}},
  \bibinfo{author}{\bibfnamefont{T.}~\bibnamefont{{Yamaguchi}}},
  \bibnamefont{and} \bibinfo{author}{\bibfnamefont{Y.}~\bibnamefont{{Takano}}},
  \bibinfo{journal}{Appl. Phys. Lett.} \textbf{\bibinfo{volume}{93}},
  \bibinfo{pages}{152505} (\bibinfo{year}{2008}).

\bibitem[{\citenamefont{Garbarino et~al.}(2009)\citenamefont{Garbarino, Sow,
  Lejay, Sulpice, Toulemonde, Mezouar, and
  N{\'u}{\~n}ez-Regueiro}}]{garbarino2009EPL}
\bibinfo{author}{\bibfnamefont{G.}~\bibnamefont{Garbarino}},
  \bibinfo{author}{\bibfnamefont{A.}~\bibnamefont{Sow}},
  \bibinfo{author}{\bibfnamefont{P.}~\bibnamefont{Lejay}},
  \bibinfo{author}{\bibfnamefont{A.}~\bibnamefont{Sulpice}},
  \bibinfo{author}{\bibfnamefont{P.}~\bibnamefont{Toulemonde}},
  \bibinfo{author}{\bibfnamefont{M.}~\bibnamefont{Mezouar}}, \bibnamefont{and}
  \bibinfo{author}{\bibfnamefont{M.}~\bibnamefont{N{\'u}{\~n}ez-Regueiro}},
  \bibinfo{journal}{Europhys. Lett.} \textbf{\bibinfo{volume}{86}},
  \bibinfo{pages}{27001} (\bibinfo{year}{2009}).

\bibitem[{\citenamefont{Margadonna et~al.}(2009)\citenamefont{Margadonna,
  Takabayashi, Ohishi, Mizuguchi, Takano, Kagayama, Nakagawa, Takata, and
  Prassides}}]{margadonna2009PRB}
\bibinfo{author}{\bibfnamefont{S.}~\bibnamefont{Margadonna}},
  \bibinfo{author}{\bibfnamefont{Y.}~\bibnamefont{Takabayashi}},
  \bibinfo{author}{\bibfnamefont{Y.}~\bibnamefont{Ohishi}},
  \bibinfo{author}{\bibfnamefont{Y.}~\bibnamefont{Mizuguchi}},
  \bibinfo{author}{\bibfnamefont{Y.}~\bibnamefont{Takano}},
  \bibinfo{author}{\bibfnamefont{T.}~\bibnamefont{Kagayama}},
  \bibinfo{author}{\bibfnamefont{T.}~\bibnamefont{Nakagawa}},
  \bibinfo{author}{\bibfnamefont{M.}~\bibnamefont{Takata}}, \bibnamefont{and}
  \bibinfo{author}{\bibfnamefont{K.}~\bibnamefont{Prassides}},
  \bibinfo{journal}{Phys. Rev. B} \textbf{\bibinfo{volume}{80}},
  \bibinfo{pages}{064506} (\bibinfo{year}{2009}).

\bibitem[{\citenamefont{Yeh et~al.}(2008)\citenamefont{Yeh, Huang, Huang, Chen,
  Hsu, Wu, Lee, Chu, Chen, Luo et~al.}}]{yeh2008}
\bibinfo{author}{\bibfnamefont{K.-W.} \bibnamefont{Yeh}},
  \bibinfo{author}{\bibfnamefont{T.-W.} \bibnamefont{Huang}},
  \bibinfo{author}{\bibfnamefont{Y.-L.} \bibnamefont{Huang}},
  \bibinfo{author}{\bibfnamefont{T.-K.} \bibnamefont{Chen}},
  \bibinfo{author}{\bibfnamefont{F.-C.} \bibnamefont{Hsu}},
  \bibinfo{author}{\bibfnamefont{P.~M.} \bibnamefont{Wu}},
  \bibinfo{author}{\bibfnamefont{Y.-C.} \bibnamefont{Lee}},
  \bibinfo{author}{\bibfnamefont{Y.-Y.} \bibnamefont{Chu}},
  \bibinfo{author}{\bibfnamefont{C.-L.} \bibnamefont{Chen}},
  \bibinfo{author}{\bibfnamefont{J.-Y.} \bibnamefont{Luo}},
  \bibnamefont{et~al.}, \bibinfo{journal}{Europhys. Lett.}
  \textbf{\bibinfo{volume}{84}}, \bibinfo{pages}{37002} (\bibinfo{year}{2008}).

\bibitem[{\citenamefont{{Yoshida} et~al.}(2009)\citenamefont{{Yoshida},
  {Wakita}, {Okazaki}, {Mizuguchi}, {Tsuda}, {Takano}, {Takeya}, {Hirata},
  {Muro}, {Okawa} et~al.}}]{yoshida2009JPSJ}
\bibinfo{author}{\bibfnamefont{R.}~\bibnamefont{{Yoshida}}},
  \bibinfo{author}{\bibfnamefont{T.}~\bibnamefont{{Wakita}}},
  \bibinfo{author}{\bibfnamefont{H.}~\bibnamefont{{Okazaki}}},
  \bibinfo{author}{\bibfnamefont{Y.}~\bibnamefont{{Mizuguchi}}},
  \bibinfo{author}{\bibfnamefont{S.}~\bibnamefont{{Tsuda}}},
  \bibinfo{author}{\bibfnamefont{Y.}~\bibnamefont{{Takano}}},
  \bibinfo{author}{\bibfnamefont{H.}~\bibnamefont{{Takeya}}},
  \bibinfo{author}{\bibfnamefont{K.}~\bibnamefont{{Hirata}}},
  \bibinfo{author}{\bibfnamefont{T.}~\bibnamefont{{Muro}}},
  \bibinfo{author}{\bibfnamefont{M.}~\bibnamefont{{Okawa}}},
  \bibnamefont{et~al.}, \bibinfo{journal}{J. Phys. Soc. Jpn.}
  \textbf{\bibinfo{volume}{78}}, \bibinfo{pages}{034708}
  (\bibinfo{year}{2009}).

\bibitem[{\citenamefont{{Yamasaki} et~al.}()\citenamefont{{Yamasaki}, {Imada},
  {Takase}, {Muro}, {Kato}, {Kobori}, {Sugimura}, {Umeyama}, {Sato}, {Hara}
  et~al.}}]{yamasaki2009u}
\bibinfo{author}{\bibfnamefont{A.}~\bibnamefont{{Yamasaki}}},
  \bibinfo{author}{\bibfnamefont{S.}~\bibnamefont{{Imada}}},
  \bibinfo{author}{\bibfnamefont{K.}~\bibnamefont{{Takase}}},
  \bibinfo{author}{\bibfnamefont{T.}~\bibnamefont{{Muro}}},
  \bibinfo{author}{\bibfnamefont{Y.}~\bibnamefont{{Kato}}},
  \bibinfo{author}{\bibfnamefont{H.}~\bibnamefont{{Kobori}}},
  \bibinfo{author}{\bibfnamefont{A.}~\bibnamefont{{Sugimura}}},
  \bibinfo{author}{\bibfnamefont{N.}~\bibnamefont{{Umeyama}}},
  \bibinfo{author}{\bibfnamefont{H.}~\bibnamefont{{Sato}}},
  \bibinfo{author}{\bibfnamefont{Y.}~\bibnamefont{{Hara}}},
  \bibnamefont{et~al.}, \bibinfo{note}{arXiv/0902.3314}.

\bibitem[{\citenamefont{{Xia} et~al.}(2009)\citenamefont{{Xia}, {Qian}, {Wray},
  {Hsieh}, {Chen}, {Luo}, {Wang}, and {Hasan}}}]{xia2009}
\bibinfo{author}{\bibfnamefont{Y.}~\bibnamefont{{Xia}}},
  \bibinfo{author}{\bibfnamefont{D.}~\bibnamefont{{Qian}}},
  \bibinfo{author}{\bibfnamefont{L.}~\bibnamefont{{Wray}}},
  \bibinfo{author}{\bibfnamefont{D.}~\bibnamefont{{Hsieh}}},
  \bibinfo{author}{\bibfnamefont{G.~F.} \bibnamefont{{Chen}}},
  \bibinfo{author}{\bibfnamefont{J.~L.} \bibnamefont{{Luo}}},
  \bibinfo{author}{\bibfnamefont{N.~L.} \bibnamefont{{Wang}}},
  \bibnamefont{and} \bibinfo{author}{\bibfnamefont{M.~Z.}
  \bibnamefont{{Hasan}}}, \bibinfo{journal}{Phys. Rev. Lett.}
  \textbf{\bibinfo{volume}{103}}, \bibinfo{pages}{037002}
  (\bibinfo{year}{2009}).

\bibitem[{\citenamefont{Nakayama et~al.}()\citenamefont{Nakayama, Sato,
  Richard, Kawahara, Sekiba, Qian, Chen, Luo, Wang, Ding
  et~al.}}]{nakayama2009u}
\bibinfo{author}{\bibfnamefont{K.}~\bibnamefont{Nakayama}},
  \bibinfo{author}{\bibfnamefont{T.}~\bibnamefont{Sato}},
  \bibinfo{author}{\bibfnamefont{P.}~\bibnamefont{Richard}},
  \bibinfo{author}{\bibfnamefont{T.}~\bibnamefont{Kawahara}},
  \bibinfo{author}{\bibfnamefont{Y.}~\bibnamefont{Sekiba}},
  \bibinfo{author}{\bibfnamefont{T.}~\bibnamefont{Qian}},
  \bibinfo{author}{\bibfnamefont{G.~F.} \bibnamefont{Chen}},
  \bibinfo{author}{\bibfnamefont{J.~L.} \bibnamefont{Luo}},
  \bibinfo{author}{\bibfnamefont{N.~L.} \bibnamefont{Wang}},
  \bibinfo{author}{\bibfnamefont{H.}~\bibnamefont{Ding}}, \bibnamefont{et~al.},
  \bibinfo{note}{arXiv/0907.0763}.

\bibitem[{\citenamefont{{Tamai} et~al.}(2010)\citenamefont{{Tamai}, {Ganin},
  {Rozbicki}, {Bacsa}, {Meevasana}, {King}, {Caffio}, {Schaub}, {Margadonna},
  {Prassides} et~al.}}]{tamai2010PRL}
\bibinfo{author}{\bibfnamefont{A.}~\bibnamefont{{Tamai}}},
  \bibinfo{author}{\bibfnamefont{A.~Y.} \bibnamefont{{Ganin}}},
  \bibinfo{author}{\bibfnamefont{E.}~\bibnamefont{{Rozbicki}}},
  \bibinfo{author}{\bibfnamefont{J.}~\bibnamefont{{Bacsa}}},
  \bibinfo{author}{\bibfnamefont{W.}~\bibnamefont{{Meevasana}}},
  \bibinfo{author}{\bibfnamefont{P.~D.~C.} \bibnamefont{{King}}},
  \bibinfo{author}{\bibfnamefont{M.}~\bibnamefont{{Caffio}}},
  \bibinfo{author}{\bibfnamefont{R.}~\bibnamefont{{Schaub}}},
  \bibinfo{author}{\bibfnamefont{S.}~\bibnamefont{{Margadonna}}},
  \bibinfo{author}{\bibfnamefont{K.}~\bibnamefont{{Prassides}}},
  \bibnamefont{et~al.}, \bibinfo{journal}{Phys. Rev. Lett.}
  \textbf{\bibinfo{volume}{104}} (\bibinfo{year}{2010}).

\bibitem[{\citenamefont{{Sales} et~al.}(2009)\citenamefont{{Sales}, {Sefat},
  {McGuire}, {Jin}, {Mandrus}, and {Mozharivskyj}}}]{sales2009PRB}
\bibinfo{author}{\bibfnamefont{B.~C.} \bibnamefont{{Sales}}},
  \bibinfo{author}{\bibfnamefont{A.~S.} \bibnamefont{{Sefat}}},
  \bibinfo{author}{\bibfnamefont{M.~A.} \bibnamefont{{McGuire}}},
  \bibinfo{author}{\bibfnamefont{R.~Y.} \bibnamefont{{Jin}}},
  \bibinfo{author}{\bibfnamefont{D.}~\bibnamefont{{Mandrus}}},
  \bibnamefont{and}
  \bibinfo{author}{\bibfnamefont{Y.}~\bibnamefont{{Mozharivskyj}}},
  \bibinfo{journal}{Phys. Rev. B} \textbf{\bibinfo{volume}{79}},
  \bibinfo{pages}{094521} (\bibinfo{year}{2009}).

\bibitem[{\citenamefont{{Subedi} et~al.}(2008)\citenamefont{{Subedi}, {Zhang},
  {Singh}, and {Du}}}]{subedi2008PRB}
\bibinfo{author}{\bibfnamefont{A.}~\bibnamefont{{Subedi}}},
  \bibinfo{author}{\bibfnamefont{L.}~\bibnamefont{{Zhang}}},
  \bibinfo{author}{\bibfnamefont{D.~J.} \bibnamefont{{Singh}}},
  \bibnamefont{and} \bibinfo{author}{\bibfnamefont{M.~H.} \bibnamefont{{Du}}},
  \bibinfo{journal}{Phys. Rev. B} \textbf{\bibinfo{volume}{78}},
  \bibinfo{pages}{134514} (\bibinfo{year}{2008}).

\bibitem[{\citenamefont{{Miyake} et~al.}(2010)\citenamefont{{Miyake},
  {Nakamura}, {Arita}, and {Imada}}}]{miyake2010}
\bibinfo{author}{\bibfnamefont{T.}~\bibnamefont{{Miyake}}},
  \bibinfo{author}{\bibfnamefont{K.}~\bibnamefont{{Nakamura}}},
  \bibinfo{author}{\bibfnamefont{R.}~\bibnamefont{{Arita}}}, \bibnamefont{and}
  \bibinfo{author}{\bibfnamefont{M.}~\bibnamefont{{Imada}}},
  \bibinfo{journal}{J. Phys. Soc. Jpn.} \textbf{\bibinfo{volume}{79}}
  \bibinfo{pages}{044705} (\bibinfo{year}{2010}).

\bibitem[{\citenamefont{Craco et~al.}()\citenamefont{Craco, Laad, and
  Leoni}}]{craco_2009u}
\bibinfo{author}{\bibfnamefont{L.}~\bibnamefont{Craco}},
  \bibinfo{author}{\bibfnamefont{M.~S.} \bibnamefont{Laad}}, \bibnamefont{and}
  \bibinfo{author}{\bibfnamefont{S.}~\bibnamefont{Leoni}},
  \bibinfo{note}{arXiv/0910.3828}.

\bibitem[{\citenamefont{{Werner} et~al.}(2008)\citenamefont{{Werner}, {Gull},
  {Troyer}, and {Millis}}}]{werner3band2008}
\bibinfo{author}{\bibfnamefont{P.}~\bibnamefont{{Werner}}},
  \bibinfo{author}{\bibfnamefont{E.}~\bibnamefont{{Gull}}},
  \bibinfo{author}{\bibfnamefont{M.}~\bibnamefont{{Troyer}}}, \bibnamefont{and}
  \bibinfo{author}{\bibfnamefont{A.~J.} \bibnamefont{{Millis}}},
  \bibinfo{journal}{Phys. Rev. Lett.} \textbf{\bibinfo{volume}{101}},
  \bibinfo{pages}{166405} (\bibinfo{year}{2008}).

\bibitem[{\citenamefont{Haule and Kotliar}(2009)}]{haule2}
\bibinfo{author}{\bibfnamefont{K.}~\bibnamefont{Haule}} \bibnamefont{and}
  \bibinfo{author}{\bibfnamefont{G.}~\bibnamefont{Kotliar}},
  \bibinfo{journal}{New J. Phys.} \textbf{\bibinfo{volume}{11}},
  \bibinfo{pages}{025021} (\bibinfo{year}{2009}).

\bibitem[{\citenamefont{{Ishida} and {Liebsch}}(2010)}]{ishida2010PRB}
\bibinfo{author}{\bibfnamefont{H.}~\bibnamefont{{Ishida}}} \bibnamefont{and}
  \bibinfo{author}{\bibfnamefont{A.}~\bibnamefont{{Liebsch}}},
  \bibinfo{journal}{Phys. Rev. B} \textbf{\bibinfo{volume}{81}}
  (\bibinfo{year}{2010}).

\bibitem[{\citenamefont{{Aichhorn} et~al.}(2009)\citenamefont{{Aichhorn},
  {Pourovskii}, {Vildosola}, {Ferrero}, {Parcollet}, {Miyake}, {Georges}, and
  {Biermann}}}]{aichhorn2009}
\bibinfo{author}{\bibfnamefont{M.}~\bibnamefont{{Aichhorn}}},
  \bibinfo{author}{\bibfnamefont{L.}~\bibnamefont{{Pourovskii}}},
  \bibinfo{author}{\bibfnamefont{V.}~\bibnamefont{{Vildosola}}},
  \bibinfo{author}{\bibfnamefont{M.}~\bibnamefont{{Ferrero}}},
  \bibinfo{author}{\bibfnamefont{O.}~\bibnamefont{{Parcollet}}},
  \bibinfo{author}{\bibfnamefont{T.}~\bibnamefont{{Miyake}}},
  \bibinfo{author}{\bibfnamefont{A.}~\bibnamefont{{Georges}}},
  \bibnamefont{and}
  \bibinfo{author}{\bibfnamefont{S.}~\bibnamefont{{Biermann}}},
  \bibinfo{journal}{Phys. Rev. B} \textbf{\bibinfo{volume}{80}},
  \bibinfo{pages}{085101} (\bibinfo{year}{2009}).

\bibitem[{\citenamefont{Blaha et~al.}(2001)\citenamefont{Blaha, Schwarz,
  Madsen, Kvasnicka, and Luitz}}]{Wien2k}
\bibinfo{author}{\bibfnamefont{P.}~\bibnamefont{Blaha}},
  \bibinfo{author}{\bibfnamefont{K.}~\bibnamefont{Schwarz}},
  \bibinfo{author}{\bibfnamefont{G.}~\bibnamefont{Madsen}},
  \bibinfo{author}{\bibfnamefont{D.}~\bibnamefont{Kvasnicka}},
  \bibnamefont{and} \bibinfo{author}{\bibfnamefont{J.}~\bibnamefont{Luitz}},
  \emph{\bibinfo{title}{WIEN2k, An augmented Plane Wave + Local Orbitals
  Program for Calculating Crystal Properties}} (\bibinfo{publisher}{Techn.
  Universitat Wien, Austria, ISBN 3-9501031-1-2.}, \bibinfo{year}{2001}).

\bibitem[{\citenamefont{{Margadonna} et~al.}(2008)\citenamefont{{Margadonna},
  {Takabayashi}, {McDonald}, {Kasperkiewicz}, {Mizuguchi}, {Takano}, {Fitch},
  {Suard}, and {Prassides}}}]{margadonna2008}
\bibinfo{author}{\bibfnamefont{S.}~\bibnamefont{{Margadonna}}},
  \bibinfo{author}{\bibfnamefont{Y.}~\bibnamefont{{Takabayashi}}},
  \bibinfo{author}{\bibfnamefont{M.~T.} \bibnamefont{{McDonald}}},
  \bibinfo{author}{\bibfnamefont{K.}~\bibnamefont{{Kasperkiewicz}}},
  \bibinfo{author}{\bibfnamefont{Y.}~\bibnamefont{{Mizuguchi}}},
  \bibinfo{author}{\bibfnamefont{Y.}~\bibnamefont{{Takano}}},
  \bibinfo{author}{\bibfnamefont{A.~N.} \bibnamefont{{Fitch}}},
  \bibinfo{author}{\bibfnamefont{E.}~\bibnamefont{{Suard}}}, \bibnamefont{and}
  \bibinfo{author}{\bibfnamefont{K.}~\bibnamefont{{Prassides}}},
  \bibinfo{journal}{Chem. Commun.} \textbf{\bibinfo{volume}{43}},
  \bibinfo{pages}{5607} (\bibinfo{year}{2008}).

\bibitem[{\citenamefont{Werner et~al.}(2006)\citenamefont{Werner, Comanac, de'
  Medici, Troyer, and Millis}}]{werner_ctqmc}
\bibinfo{author}{\bibfnamefont{P.}~\bibnamefont{Werner}},
  \bibinfo{author}{\bibfnamefont{A.}~\bibnamefont{Comanac}},
  \bibinfo{author}{\bibfnamefont{L.}~\bibnamefont{de' Medici}},
  \bibinfo{author}{\bibfnamefont{M.}~\bibnamefont{Troyer}}, \bibnamefont{and}
  \bibinfo{author}{\bibfnamefont{A.~J.} \bibnamefont{Millis}},
  \bibinfo{journal}{Phys. Rev. Lett.} \textbf{\bibinfo{volume}{97}},
  \bibinfo{eid}{076405} (\bibinfo{year}{2006}).

\bibitem[{\citenamefont{Aryasetiawan et~al.}(2004)\citenamefont{Aryasetiawan,
  Imada, Georges, Kotliar, Biermann, and Lichtenstein}}]{ferdi_cRPA}
\bibinfo{author}{\bibfnamefont{F.}~\bibnamefont{Aryasetiawan}},
  \bibinfo{author}{\bibfnamefont{M.}~\bibnamefont{Imada}},
  \bibinfo{author}{\bibfnamefont{A.}~\bibnamefont{Georges}},
  \bibinfo{author}{\bibfnamefont{G.}~\bibnamefont{Kotliar}},
  \bibinfo{author}{\bibfnamefont{S.}~\bibnamefont{Biermann}}, \bibnamefont{and}
  \bibinfo{author}{\bibfnamefont{A.~I.} \bibnamefont{Lichtenstein}},
  \bibinfo{journal}{Phys. Rev. B} \textbf{\bibinfo{volume}{70}},
  \bibinfo{pages}{195104} (\bibinfo{year}{2004}).

\bibitem[{\citenamefont{{Miyake} et~al.}(2008)\citenamefont{{Miyake},
  {Pourovskii}, {Vildosola}, {Biermann}, and {Georges}}}]{miyake1}
\bibinfo{author}{\bibfnamefont{T.}~\bibnamefont{{Miyake}}},
  \bibinfo{author}{\bibfnamefont{L.}~\bibnamefont{{Pourovskii}}},
  \bibinfo{author}{\bibfnamefont{V.}~\bibnamefont{{Vildosola}}},
  \bibinfo{author}{\bibfnamefont{S.}~\bibnamefont{{Biermann}}},
  \bibnamefont{and}
  \bibinfo{author}{\bibfnamefont{A.}~\bibnamefont{{Georges}}},
  \bibinfo{journal}{J. Phys. Soc. Jpn.} \textbf{\bibinfo{volume}{77}},
  \bibinfo{pages}{99} (\bibinfo{year}{2008}).

\bibitem[{\citenamefont{Ylvisaker et~al.}(2009)\citenamefont{Ylvisaker,
  Pickett, and Koepernik}}]{Ylvisaker_LSDA+U_2009_prb}
\bibinfo{author}{\bibfnamefont{E.~R.} \bibnamefont{Ylvisaker}},
  \bibinfo{author}{\bibfnamefont{W.~E.} \bibnamefont{Pickett}},
  \bibnamefont{and}
  \bibinfo{author}{\bibfnamefont{K.}~\bibnamefont{Koepernik}},
  \bibinfo{journal}{Phys. Rev. B} \textbf{\bibinfo{volume}{79}},
  \bibinfo{pages}{035103} (\bibinfo{year}{2009}).

\bibitem[{\citenamefont{{Beach}}()}]{beach_ME}
\bibinfo{author}{\bibfnamefont{K.~S.~D.} \bibnamefont{{Beach}}},
  \bibinfo{note}{arXiv/0403055}.

\bibitem[{\citenamefont{{Malaeb} et~al.}(2008)\citenamefont{{Malaeb},
  {Yoshida}, {Kataoka}, {Fujimori}, {Kubota}, {Ono}, {Usui}, {Kuroki}, {Arita},
  {Aoki} et~al.}}]{Malaeb1}
\bibinfo{author}{\bibfnamefont{W.}~\bibnamefont{{Malaeb}}},
  \bibinfo{author}{\bibfnamefont{T.}~\bibnamefont{{Yoshida}}},
  \bibinfo{author}{\bibfnamefont{T.}~\bibnamefont{{Kataoka}}},
  \bibinfo{author}{\bibfnamefont{A.}~\bibnamefont{{Fujimori}}},
  \bibinfo{author}{\bibfnamefont{M.}~\bibnamefont{{Kubota}}},
  \bibinfo{author}{\bibfnamefont{K.}~\bibnamefont{{Ono}}},
  \bibinfo{author}{\bibfnamefont{H.}~\bibnamefont{{Usui}}},
  \bibinfo{author}{\bibfnamefont{K.}~\bibnamefont{{Kuroki}}},
  \bibinfo{author}{\bibfnamefont{R.}~\bibnamefont{{Arita}}},
  \bibinfo{author}{\bibfnamefont{H.}~\bibnamefont{{Aoki}}},
  \bibnamefont{et~al.}, \bibinfo{journal}{J. Phys. Soc. Jpn.}
  \textbf{\bibinfo{volume}{77}}, \bibinfo{pages}{093714}
  (\bibinfo{year}{2008}).

\bibitem[{\citenamefont{Pruschke and Bulla}(2005)}]{pruschke2005}
\bibinfo{author}{\bibfnamefont{T.}~\bibnamefont{Pruschke}} \bibnamefont{and}
  \bibinfo{author}{\bibfnamefont{R.}~\bibnamefont{Bulla}},
  \bibinfo{journal}{Eur. Phys. J. B} \textbf{\bibinfo{volume}{44}},
  \bibinfo{pages}{217} (\bibinfo{year}{2005}).

\bibitem[{\citenamefont{{de Medici} et~al.}(2009)\citenamefont{{de Medici},
  Hassan, and Capone}}]{medici2009}
\bibinfo{author}{\bibfnamefont{L.}~\bibnamefont{{de Medici}}},
  \bibinfo{author}{\bibfnamefont{S.~R.} \bibnamefont{Hassan}},
  \bibnamefont{and} \bibinfo{author}{\bibfnamefont{M.}~\bibnamefont{Capone}},
  \bibinfo{journal}{J. Supercond. Nov. Magn.} \textbf{\bibinfo{volume}{22}},
  \bibinfo{pages}{535} (\bibinfo{year}{2009}).

\bibitem[{\citenamefont{Liebsch and Ishida}()}]{liebsch2010u}
\bibinfo{author}{\bibfnamefont{A.}~\bibnamefont{Liebsch}} \bibnamefont{and}
  \bibinfo{author}{\bibfnamefont{H.}~\bibnamefont{Ishida}},
  \bibinfo{note}{arxiv/1004.2851}.

\bibitem[{\citenamefont{Braithwaite et~al.}(2009)\citenamefont{Braithwaite,
  Salce, Lapertot, Bourdarot, Marin, Aoki, and Hanfland}}]{braithwaite2009}
\bibinfo{author}{\bibfnamefont{D.}~\bibnamefont{Braithwaite}},
  \bibinfo{author}{\bibfnamefont{B.}~\bibnamefont{Salce}},
  \bibinfo{author}{\bibfnamefont{G.}~\bibnamefont{Lapertot}},
  \bibinfo{author}{\bibfnamefont{F.}~\bibnamefont{Bourdarot}},
  \bibinfo{author}{\bibfnamefont{C.}~\bibnamefont{Marin}},
  \bibinfo{author}{\bibfnamefont{D.}~\bibnamefont{Aoki}}, \bibnamefont{and}
  \bibinfo{author}{\bibfnamefont{M.}~\bibnamefont{Hanfland}},
  \bibinfo{journal}{J. Phys. Cond. Mat.} \textbf{\bibinfo{volume}{21}},
  \bibinfo{pages}{232202} (\bibinfo{year}{2009}).

\bibitem[{\citenamefont{Imai et~al.}(2009)\citenamefont{Imai, Ahilan, Ning,
  McQueen, and Cava}}]{imai_PRL2009}
\bibinfo{author}{\bibfnamefont{T.}~\bibnamefont{Imai}},
  \bibinfo{author}{\bibfnamefont{K.}~\bibnamefont{Ahilan}},
  \bibinfo{author}{\bibfnamefont{F.~L.} \bibnamefont{Ning}},
  \bibinfo{author}{\bibfnamefont{T.~M.} \bibnamefont{McQueen}},
  \bibnamefont{and} \bibinfo{author}{\bibfnamefont{R.~J.} \bibnamefont{Cava}},
  \bibinfo{journal}{Phys. Rev. Lett.} \textbf{\bibinfo{volume}{102}},
  \bibinfo{pages}{177005} (\bibinfo{year}{2009}).

\bibitem[{\citenamefont{Lumsden et~al.}(2010)\citenamefont{Lumsden,
  Christianson, Goremychkin, Nagler, Mook, Stone, Abernathy, Guidi, MacDougall,
  de~la Cruz et~al.}}]{lumsden_NaturePhysics2010}
\bibinfo{author}{\bibfnamefont{M.~D.} \bibnamefont{Lumsden}},
  \bibinfo{author}{\bibfnamefont{A.~D.} \bibnamefont{Christianson}},
  \bibinfo{author}{\bibfnamefont{E.~A.} \bibnamefont{Goremychkin}},
  \bibinfo{author}{\bibfnamefont{S.~E.} \bibnamefont{Nagler}},
  \bibinfo{author}{\bibfnamefont{H.~A.} \bibnamefont{Mook}},
  \bibinfo{author}{\bibfnamefont{M.~B.} \bibnamefont{Stone}},
  \bibinfo{author}{\bibfnamefont{D.~L.} \bibnamefont{Abernathy}},
  \bibinfo{author}{\bibfnamefont{T.}~\bibnamefont{Guidi}},
  \bibinfo{author}{\bibfnamefont{G.~J.} \bibnamefont{MacDougall}},
  \bibinfo{author}{\bibfnamefont{C.}~\bibnamefont{de~la Cruz}},
  \bibnamefont{et~al.}, \bibinfo{journal}{Nature Physics}
  \textbf{\bibinfo{volume}{6}}, \bibinfo{pages}{182} (\bibinfo{year}{2010}).

\end{thebibliography}
\end{document}